\documentstyle[12pt,epsf,epsfig]{article}
\begin{document}
\baselineskip 0.6cm
\newcommand{\gsim}{ \mathop{}_{\textstyle \sim}^{\textstyle >} }
\newcommand{\lsim}{ \mathop{}_{\textstyle \sim}^{\textstyle <} }
\newcommand{\vev}[1]{ \langle {#1} \rangle }
\newcommand{\EV}{ {\rm eV} }
\newcommand{\KEV}{ {\rm keV} }
\newcommand{\MEV}{ {\rm MeV} }
\newcommand{\GEV}{ {\rm GeV} }
\newcommand{\TEV}{ {\rm TeV} }
\newcommand{\eps}{\varepsilon}
\newcommand{\barr}[1]{ \overline{{#1}} }
\newcommand{\del}{\partial}
\newcommand{\nn}{\nonumber}
\newcommand{\ra}{\rightarrow}
\newcommand{\bino}{\tilde{\chi}}
\def\tr{\mathop{\rm tr}\nolimits}
\def\Tr{\mathop{\rm Tr}\nolimits}
\def\Re{\mathop{\rm Re}\nolimits}
\def\Im{\mathop{\rm Im}\nolimits}
\setcounter{footnote}{1}

\begin{titlepage}

\begin{flushright}
UT-02-11\\
\end{flushright}

\vskip 3cm
\begin{center}
{\Large \bf Leptogenesis with Almost Degenerate Majorana Neutrinos}
\vskip 2.4cm

\center{Masaaki~Fujii$^{1}$, K.~Hamaguchi$^{1}$ and T.~Yanagida$^{1,2}$}\\
$^{1}${\it{Department of Physics, University of Tokyo, Tokyo 113-0033,
Japan}}\\
$^{2}${\it{Research Center for the Early Universe, University of
Tokyo, Tokyo, 113-0033, Japan}}

\vskip 3cm

\abstract{We investigate the leptogenesis with almost degenerate
 neutrinos, in the framework of democratic mass matrix, which naturally
 explains the large mixing angles for neutrino oscillations as well as
 quark masses and mixing matrix. We find that the baryon asymmetry in
 the present universe is explained via the decays of right-handed
 neutrinos produced nonthermally by the inflaton decay. The model
 predicts neutrinoless double beta decays accessible in near future
 experiments.}

\end{center}
\end{titlepage}

\renewcommand{\thefootnote}{\arabic{footnote}}
\setcounter{footnote}{0}

%
%
%
%

One of the fundamental problems in particle physics is to understand the
observed masses and mixing angles of quarks and leptons. There have
been, in fact, proposed many models of the mass matrices for quarks and
leptons. Among them the hypothesis of democratic mass
matrix~\cite{Harari-etal} is very interesting not only theoretically,
but also phenomenologically. Here, the three families of quarks and
leptons are treated in an equal footing, i.e. a permutation symmetry
S$_3 \times $S$_3$ is imposed among three families. The democratic mass
matrix has been known very successful in explaining the observed masses
and mixing angles for quarks~\cite{Koide}. This approach has been
extended to the charged lepton and neutrino sectors, and it has been
found that the observed large mixings of neutrinos can be rather easily
understood due to the nearly diagonal neutrino mass
matrix~\cite{FTY,Fritzsch-etal-2}.\footnote{Ref.~\cite{Fritzsch-etal-1}
considers Dirac or Majorana neutrinos. Thus, their model does not have a
group theoretical reason for the family-diagonal and degenerate neutrino
masses.}  Surprisingly, the model also accounts for the small mixing
angle U$_{{\rm e}3}$, required from the CHOOZ experiment~\cite{CHOOZ},
although other two mixing angles are
large~\cite{SK-Atm,SK-solar}. Furthermore, this scheme yields almost
degenerate Majorana masses for neutrinos ant it predicts neutrinoless
double beta decays accessible in near future experiments, such as
GENIUS~\cite{GENIUS}, CUORE~\cite{CUORE}, MOON~\cite{MOON},
XMASS~\cite{XMASS} and EXO~\cite{EXO}.

In this letter, we discuss the leptogenesis~\cite{FY} in the framework
of the democratic mass matrix. We show that the baryon asymmetry in the
present universe can be generated via the decay of heavy right-handed
neutrinos. Here, the right-handed neutrinos are assumed to be produced
nonthermally through inflaton decay~\cite{lep-inf}.

{\it Democratic mass matrix}

Let us start by giving a brief review on the work in Ref.~\cite{FTY},
which imposes a permutation symmetry S$_3(L)\times$ S$_3(R)$ on the
three families of left-handed lepton doublets ($l_{Li}$) and
right-handed charged leptons ($e_{Ri}$). We assume that $l_{Li}$
transform as ${\bf 3}_L = {\bf 2}_L + {\bf 1}_L$ under S$_3(L)$, and
$e_{Ri}$ transform as ${\bf 3}_R = {\bf 2}_R + {\bf 1}_R$ under
S$_3(R)$.  Then the mass matrix for charged leptons which is invariant
under the S$_3(L)\times$ S$_3(R)$ is uniquely determined as follows:
\begin{eqnarray}
 \label{EQ-demo}
  M_l^{(0)}
  =
  \frac{{\cal{M}}_l}{3}
  \left[
   \begin{array}{ccc}
    1 & 1 & 1\\
    1 & 1 & 1\\
    1 & 1 & 1
   \end{array}
   \right]
   \,.
\end{eqnarray}
This mass matrix is often called ``democratic'' mass
matrix~\cite{Harari-etal}. Two mass eigenvalues of this mass matrix
vanish. Thus, we introduce a small symmetry-breaking mass term as
perturbations with a diagonal mass matrix~\cite{Koide}
\begin{eqnarray}
 M_l^{(1)}
  =
  {\cal{M}}_{l}\left[
   \begin{array}{ccc}
    -\eps_l & 0 & 0\\
    0 & \eps_l & 0\\
    0 & 0 & \delta_l
   \end{array}
   \right]
   \,.
\end{eqnarray}
Then the matrix $M_l = M_l^{(0)} + M_l^{(1)}$ is diagonalized as
\begin{eqnarray}
 V_l^{\dagger} M_l V_l = {\rm diag}(m^l_1, m^l_2, m^l_3)
  \,,
\end{eqnarray}
where
\begin{eqnarray}
 m_1^l = {\cal{M}}_{l}\left(\frac{\delta_l}{3} - \frac{\xi_l}{6}\right)
  \,,\qquad
 m_2^l = {\cal{M}}_{l}\left(\frac{\delta_l}{3} + \frac{\xi_l}{6}\right)
  \,,\qquad
 m_3^l = {\cal{M}}_{l}\left(1 + \frac{\delta_l}{3}\right)
  \,,
\end{eqnarray}
with
\begin{eqnarray}
 \xi_l = 2\left(
	   \delta_l^2 + 3 \eps_l^2
	   \right)^{1/2}
  \,,
\end{eqnarray}
where terms with higher order of $\delta_l$ and $\eps_l$ have been
ignored. The unitary matrix $V_l$ which diagonalizes the mass matrix
$M_l$ is given by $V_l = A B_l$, where
\begin{eqnarray}
 A = 
  \left[
   \begin{array}{ccc}
    1 / \sqrt{2} & 1 / \sqrt{6} & 1 / \sqrt{3}\\
    -1/\sqrt{2} & 1 / \sqrt{6} & 1 / \sqrt{3}\\
    0 & -2 / \sqrt{6} & 1 / \sqrt{3}
   \end{array}
   \right]
   \label{EQ-lepton-angle1},
\end{eqnarray}
and
\begin{eqnarray}
 B_l = 
  \left[
   \begin{array}{ccc}
    \cos\theta_l & -\sin\theta_l & -\lambda_l \sin 2\theta_l
     \\
    \sin\theta_l & \cos\theta_l & -\lambda_l \cos 2\theta_l
     \\
    \lambda_l \sin 3\theta_l & \lambda_l \cos 3\theta_l & 1
   \end{array}
   \right]
   \,,
   \quad
   \tan 2\theta_l \simeq \frac{\sqrt{3}\eps_l}{\delta_l}
   \,,
   \quad
   \lambda_l \simeq \frac{\xi_l}{3\sqrt{2}}
   \,.
   \label{EQ-lepton-angle}
\end{eqnarray}
The empirical masses for the charged leptons
($m_1^l = m_e$, $m_2^l = m_\mu$ and $m_3^l = m_\tau$) are given by
adjusting appropriately the parameters ${\cal{M}}_l$, $\eps_l$ and
$\delta_l$ as
\begin{equation}
 {\cal{M}}_{l}=1719\;[\MEV],\;\;\eps_l=0.00665,\;\;\delta_{l}=0.095\;\;,
  \label{EQ-lepton-para}
\end{equation}
which leads to $\sin\theta_l\simeq 0.066$.

Let us turn to the neutrino sector. Before introducing heavy
right-handed neutrinos, we discuss the mass matrix for the light
Majorana neutrinos in the effective non-renormalizable operator:
\begin{eqnarray}
 \frac{f_{ij}}{2 M}
  \l_{Li} H\,\,\l_{Lj} H
  \,,
\end{eqnarray}
where $H$ denotes the Higgs doublet. There are two independent mass
matrices that are invariants under S$_3(L)$ symmetry~\cite{FTY}:
\begin{eqnarray}
 \label{EQ-S3S3}
 \left[
  \begin{array}{ccc}
   1 & 0 & 0\\
   0 & 1 & 0\\
   0 & 0 & 1
  \end{array}
  \right]\,,
  \quad
 \left[
  \begin{array}{ccc}
   0 & 1 & 1\\
   1 & 0 & 1\\
   1 & 1 & 0
  \end{array}
  \right]\,.  
\label{inv-M}
\end{eqnarray}
Here we take the first form for the time being, since it automatically leads to the large
mixing angles for both of the solar and the atmospheric neutrinos. (We
will add the second mass matrix later.) Then, the neutrinos are
degenerate in mass. With symmetry breaking terms in diagonal elements,
the neutrino mass matrix is given by
\begin{eqnarray}
 \label{EQ-mass-nu}
  M_\nu =\frac{f \vev{H}^2}{M}
  =
  {\cal{M}}_\nu
  \left[
   \begin{array}{ccc}
    1 & &
     \\
     & 1 + \eps_\nu & 
     \\
     &  & 1 + \delta_\nu
   \end{array}
   \right]\,.
\label{D-Nmatrix}
\end{eqnarray}

The neutrino-mixing matrix in the basis where the mass matrix for the
charged leptons is diagonal is given by $U = (V_l)^{\dagger}$
as~\footnote{ Here, we have neglected the small mixing angle
$\theta_{l}$ of charged leptons.}
\begin{eqnarray}
 U = (A B_l)^{\dagger} \simeq A^{\dagger}
  =
  \left[
   \begin{array}{ccc}
    1 / \sqrt{2} & -1/\sqrt{2} & 0   \\
     1 / \sqrt{6} & 1 / \sqrt{6} & -2 / \sqrt{6} \\
    1 / \sqrt{3} & 1 / \sqrt{3} & 1 / \sqrt{3}
   \end{array}
   \right]
   \,.
   \label{EQ-bi-maximal}
\end{eqnarray}
As denoted, this indicates nearly bi-maximal neutrino oscillations,
i.e.,
\begin{eqnarray}
 \sin^2 2\theta_{12}\simeq 1\,,
  \qquad \sin^2 2\theta_{23}\simeq \frac{8}{9}\,,
  \label{sol-angle}
\end{eqnarray}
where $\theta_{12}$ and $\theta_{23}$ are the solar and the atmospheric
neutrino mixing angles, respectively.  Furthermore, the U$_{{\rm e}3}$
element automatically vanishes, U$_{{\rm e}3}\simeq 0$. The mass squared
differences for solar and atmospheric  neutrino oscillations are given by
$\Delta m^2_{\rm solar} \simeq 2 {\cal{M}}_\nu^2 \eps_\nu$ and $\Delta
m^2_{\rm atm} \simeq 2 {\cal{M}}_\nu^2 \delta_\nu$, and hence the
breaking parameters should satisfy $\eps_\nu/\delta_\nu\simeq \Delta
m^2_{\rm solar}/\Delta m^2_{\rm atm}$.  The experimentally observed
values for these quantities are given by~\cite{SK-Atm,postSNO}
\begin{eqnarray}
 &&\Delta m^2_{\rm{atm}}=\{(1.0-6.0),\;3.2\}\times 10^{-3} \EV^2\nonumber\\
 &&\Delta m^{2}_{\rm{solar}}=\{(2-50),\;4.9\}\times 10^{-5} \EV^2\nonumber\\
 &&{\rm{sin}}^{2}2\theta_{23}=\{(0.83-1),\;1\}\nonumber\\
 &&{\rm{tan}}^{2}\theta_{12}=\{(0.2-0.8),\;0.37\}\;,
  \label{experiment}
\end{eqnarray}
where the last numbers in the parentheses denote the best fit values.
Here, we have taken the large mixing MSW (LMA) solution to the solar
neutrino problem, which is the most preferable in the present
experiments~\cite{SK-solar}.

Recently, a positive indication for neutrinoless double beta decay has
been reported, with $\nu_e$-$\nu_e$ component of the neutrino mass
matrix $|m_{\nu_e \nu_e}|\simeq
0.05$--$0.84\EV$~\cite{H-M-0nubb}. Therefore, it might be interesting to
consider the degenerate neutrinos of mass of order ${\cal
O}(0.1\EV)$. Hereafter, we take $m_i^\nu\simeq {\cal M}_\nu\simeq
0.1\EV$ as a representation. In this case the breaking parameters in the
neutrino mass matrix are obtained as $\eps_\nu = \Delta m^2_{\rm
solar}/(2 {\cal M}_\nu^2)\simeq 0.001$--$0.025$ and $\delta_\nu = \Delta
m^2_{\rm atm}/(2 {\cal M}_\nu^2)\simeq 0.05$--$0.3$. We see that their
orders of magnitudes are just around those of the breaking parameters in
charged lepton mass matrix, $\eps_l = 0.00665$ and $\delta_l = 0.095$,
in Eq.~(\ref{EQ-lepton-para}).

One may worry about that the predicted value of the mixing angle for the
solar neutrino $\theta_{12}$ in Eq.~(\ref{sol-angle}) is too large to
fit the experimental value.  Even if we include the effect of nonzero
$\sin\theta_l \simeq 0.066$, the mixing angle of the solar neutrino
oscillation is only slightly reduced to $\sin^2 2\theta_{12}\simeq
0.99$. However, a deviation from the nearly maximal mixing for solar
neutrino oscillation can be easily implemented by introducing small
off-diagonal elements $\kappa_\nu$ in the neutrino mass matrix
Eq~(\ref{D-Nmatrix}):
\begin{eqnarray}
 M_{\nu}={\cal{M}}_{\nu}
  \left[
   \begin{array}{ccc}
    1 & \kappa_\nu & \kappa_\nu
     \\
    \kappa_\nu & 1+\eps_{\nu} & \kappa_\nu
     \\
    \kappa_\nu & \kappa_\nu & 1+\delta_{\nu}
   \end{array}
   \right]\;.
   \label{Nu-mat}
\end{eqnarray}
Notice that the matrix of the off-diagonal elements has the form of the
second matrix in Eq.~(\ref{EQ-S3S3}), which is also invariant under the
S$_3(L)$ as noted before.  The mixing matrix of neutrino-oscillations is then given by
$U=(V_{l})^{\dag}V_{\nu}$.  Here, $V_{\nu}$ is defined as
\begin{equation}
 V_{\nu}^{\dag}M_{\nu}V_{\nu}={\rm{diag}}(m_{1}^{\nu},\;m_{2}^{\nu},\;
  m_{3}^{\nu})\;,
\end{equation}
where $m_{1}^{\nu}\simeq {\cal{M}}_{\nu}$, $m_{2}^{\nu}\simeq
{\cal{M}}_{\nu}(1+\eps_{\nu})$, $m_{3}^{\nu}\simeq {\cal{M}}_{\nu}
(1+\delta_{\nu})$.  Here and hereafter, we require $\kappa_\nu <
\eps_\nu$ in order to ensure the nearly bi-large mixing for neutrino
oscillations.

The neutrino mass matrix given in Eq.~(\ref{Nu-mat}) has another
interesting prediction on the size of the U$_{e3}$.
When the neutrino mass matrix is diagonal as in Eq.~(\ref{D-Nmatrix}), 
the neutrino mixing matrix U is entirely determined by $V_{l}$
given in Eqs.~({\ref{EQ-lepton-angle1}}) and (\ref{EQ-lepton-angle}).
In this case, we can immediately obtain the U$_{e3}$ element as 
\begin{equation}
{\rm{U}}_{e3}=-\frac{2}{\sqrt{6}}\;{\rm{sin}}\;\theta_{l}+
\frac{\lambda_{l}}{\sqrt{3}}\;{\rm{sin}}\;3\theta_{l}\simeq -0.05\;.
\label{Ue3-exact}
\end{equation}
In the case of the general neutrino mass matrix, the above prediction on 
the U$_{e3}$ is shifted by the non-zero elements of the $V_{\nu}$.
The leading deviation of the U$_{e3}$ from the value presented in
Eq.~(\ref{Ue3-exact}) is given by
\begin{equation}
\Delta {\rm{U}}_{e3}\simeq \frac{{\rm{cos}}\;\theta_{l}}{\sqrt{2}}(V_{\nu 13}
-V_{\nu 23})\;.
\label{Ue3-dev}
\end{equation}

We now consider the specific case where the neutrino mass matrix is
given by Eq.~(\ref{Nu-mat}). One might naively imagine that the
deviation of the U$_{e3}$ is given by
\begin{equation}
\Delta {\rm{U}}_{e3}=
{\cal{O}}\left(\displaystyle{\frac{\kappa_{\nu}}{\delta_{\nu}}}\right)
\sim 10^{-(1-2)}
\end{equation}
from the above general argument.  If this is true, we can have no
precise prediction on the size of the U$_{e3}$. Interestingly, however,
this is not the case. Because of the symmetric form of the neutrino mass
matrix in Eq.~(\ref{Nu-mat}), the leading contribution given in
Eq.~(\ref{Ue3-dev}) beautifully cancel out.  Actually, the leading terms
in $V_{\nu 13}$ and $V_{\nu 23}$ are given by
\begin{eqnarray}
&&V_{\nu 13}= \frac{\kappa_{\nu}}{\delta_{\nu}}\;,\nonumber\\
&&V_{\nu 23}= \frac{\kappa_{\nu}}{\delta_{\nu}}+
\frac{\eps_{\nu}\kappa_{\nu}}{\delta_{\nu}^2}\;.
\label{Ue3-dev2}
\end{eqnarray}  
Hence, the deviation of the U$_{e3}$ in this model is highly
suppressed, $\Delta {\rm{U}}_{e3}\lsim 10^{-3}$, which allows us to have
a precise prediction, $|{\rm{U}}_{e3}|\simeq 0.05$.\footnote{ In fact,
$V_{\nu 13}=V_{\nu 23}$ in the limit of the vanishing $\eps_{\nu}$.  }
This will be clearly shown in the numerical calculations presented in
this letter.  In the future, this prediction on the size of the U$_{e3}$
will be tested in the long baseline experiments, such as JHF~\cite{JHF}.

The most natural way to explain small neutrino masses is the ``see-saw''
mechanism~\cite{seesaw}.  In this case, the mass matrix $M_\nu$ for the
light neutrinos is given by
\begin{equation}
M_{\nu}=m_{\nu D}^{T}M_{R}^{-1}m_{\nu D}
\end{equation}
after integrating out the heavy right-handed Majorana neutrinos
$N_{Ri}$.  Here, $M_{R}$ and $m_{\nu D}$ are the mass matrices of the
right-handed Majorana neutrinos $N_{Ri}$ and the Dirac mass for $l_{Li}$
and $N_{Rj}$, respectively.  In terms of the Yukawa couplings, the Dirac
neutrino mass matrix is given by $(m_{\nu D})_{ij}=(h_{D})_{ij}\vev{H}$,
where $(h_D)_{ij}$ is defined as ${\cal L} = (h_D)_{ij} N_{Ri} l_{Lj}
H$. The S$_3$ invariant matrices of $M_R$ and $h_D$ are uniquely
determined as
\begin{eqnarray}
 M_{R}={\cal{M}}_{R}
  \left[
   \begin{array}{ccc}
    1 & \kappa_R & \kappa_R \\
    \kappa_R &1 & \kappa_R\\
    \kappa_R & \kappa_R & 1
   \end{array}
   \right]\,,
   \quad
   h_{D}=k_{D}
   \left[
    \begin{array}{ccc}
     1 & \kappa_D & \kappa_D
      \\
     \kappa_D & 1 & \kappa_D
      \\
     \kappa_D & \kappa_D & 1
    \end{array}
    \right]\,,
\label{right-dirac-mass mat}
\end{eqnarray}
where we have assumed that the $N_{Ri}$ transform as ${\bf 3}_L = {\bf
2}_L+{\bf 1}_L$ under the S$_3(L)$ symmetry.\footnote{If $N_{Ri}$
transform as ${\bf 3}_R = {\bf 2}_R+{\bf 1}_R$ under the S$_3(R)$, the
Dirac Yukawa matrix becomes democratic one as in Eq.~(\ref{EQ-demo}),
which invalidates the almost diagonal mass matrix for light neutrinos in
Eq.~(\ref{Nu-mat}).} One can easily imagine that the assumption of
$\kappa_R \sim \kappa_D (\sim \kappa_\nu)$ for the off-diagonal elements
is the simplest and natural way to obtain the mass matrix for the light
neutrinos in Eq~(\ref{Nu-mat}) with $\eps_\nu = \delta_\nu = 0$. In
order to induce the symmetry breaking parameters $\eps_\nu$ and
$\delta_\nu$ in the neutrino mass matrix in Eq.~(\ref{Nu-mat}), we
introduce breaking parameters $\eps_D$ and $\delta_D$ in diagonal
elements of Yukawa matrix $h_D$, as in charged leptons:
\begin{eqnarray}
 h_{D}=k_{D}
  \left[
   \begin{array}{ccc}
    1 & \kappa_D & \kappa_D
     \\
    \kappa_D & 1+\eps_{D} & \kappa_D
     \\
    \kappa_D & \kappa_D & 1+\delta_{D}
   \end{array}
   \right]\,.
\end{eqnarray}
Then, we can obtain the required mass matrix of the light neutrinos by
taking the following pattern of the perturbations:
\begin{eqnarray}
 \kappa_R
  \sim 
  \kappa_D\,
  (\sim \kappa_\nu)
  \,\,<\,\, 
   \eps_D\, 
   (\sim \eps_\nu) \sim {\cal O}(0.01)
   \,\,<\,\,
    \delta_D\,
    (\sim \delta_\nu) \sim {\cal O}(0.1)\,.  
     \label{perturbation}
\end{eqnarray}

The above breakings of the S$_3$ symmetries might be understood as
follows. Suppose that the S$_3$ breakings are originated only from the
Yukawa coupling of Higgs field $H$, as in mass matrices for quarks and
charged leptons. Then, the S$_3(L)$ symmetry is broken in the neutrino
Dirac Yukawa matrix by the parameters $\eps_D$ and $\delta_D$
as in charged lepton sector, since it comes from the coupling of  Higgs
field. On the other hand, the mass matrix of the right-handed neutrinos
$M_R$ in Eq.~(\ref{right-dirac-mass mat}) is assumed to be S$_3$
invariant since it is decoupled from the Higgs field $H$.

The mass matrix $M_R$ for the heavy right-handed neutrino leads to two
exactly degenerate right-handed neutrinos and a slightly heavier/lighter
one, with masses $M_{Ri} = \{ {\cal M}_R(1-\kappa_R), {\cal
M}_R(1-\kappa_R), {\cal M}_R(1+2\kappa_R)\}$. As we will see in the
remainder of this letter, the off-diagonal elements $\kappa_R$ and
$\kappa_D$, which are required to explain the neutrino oscillation
experiments, play a crucial role in the leptogenesis.

\clearpage

{\it leptogenesis}~\cite{FY}

Now let us discuss the leptogenesis in the present model.  Notice that
the conventional leptogenesis scenario where the right-handed neutrinos
are produced by thermal scatterings after the inflation is somewhat
difficult in the case of the degenerate neutrinos of $m_1^\nu\simeq
m_2^\nu\simeq m_3^\nu\simeq {\cal O}(0.1)\EV$. This is because the
out-of-equilibrium condition cannot be satisfied in this case and the
amount of produced lepton asymmetry is strongly
suppressed~\cite{Buchmuller-Plumacher}. (See Appendix.) Therefore, we
consider the leptogenesis via decays of the right-handed neutrinos
$N_{Ri}$ which are produced non-thermally. A natural mechanism of such a
non-thermal production of $N_{Ri}$ is the decays of the inflaton
$\varphi$ into the $N_{Ri}$~\cite{lep-inf}.  Hereafter, we will consider
the supersymmetry (SUSY) theory, although the following discussion does
not change much in the non-supersymmetric case.

If $CP$ is not conserved in the Yukawa matrix $h_D$, the interference
between decay amplitudes of tree and one-loop diagrams results in the
lepton-number production~\cite{FY}. The lepton-number asymmetry per a
decay of right-handed neutrino $N_{Ri}$ is given by~\cite{FY,epsilon1}
\begin{eqnarray}
 \epsilon_i
  &\equiv&
  \frac
  {\sum_j\Gamma (N_{Ri}\to l_{Lj} + H) 
  - 
  \sum_j\Gamma (N_{Ri}\to \overline{l_{Lj}} + \overline{H})}
  {\sum_j\Gamma (N_{Ri}\to l_{Lj} + H) 
  +
  \sum_j\Gamma (N_{Ri}\to \overline{l_{Lj}} + \overline{H})}
  \nonumber\\
 &=&
  -\frac{1}{8\pi}
  \frac{1}{(h_D h_D^{\dagger})_{ii}}
  \sum_{k\ne i}
  \Im
  \left[
   \{
   \left(
    h_D h_D^{\dagger}
    \right)_{ik}
    \}^2
   \right]
   \left[
    {\cal{F}}_{V}\left(\frac{M_k^2}{M_i^2}\right)
    +
    {\cal{F}}_{S}\left(\frac{M_k^2}{M_i^2}\right)
    \right]\,,
    \label{EQ-epsilon_i}
\end{eqnarray}
where $N_{Ri}$, $l_{Lj}$, and $H$ ($\overline{l_{Lj}}$ and
$\overline{H}$) symbolically denote fermionic or scalar components of
corresponding supermultiplets (and their anti-particles), and
${\cal{F}}_{V}(x)$ and ${\cal{F}}_{S}(x)$ represent the contributions
from vertex and self-energy diagrams, respectively. In the case of the
SUSY theory, they are given by~\cite{epsilon1-SUSY}
\begin{eqnarray}
 {\cal{F}}_{V}(x) = \sqrt{x}\ln\left( 1 + \frac{1}{x}\right)
  \,,
  \qquad
  {\cal{F}}_{S}(x) = \frac{2\sqrt{x}}{x - 1}
  \,.
  \label{EQ-fs}
\end{eqnarray}
Here, we have assumed that the mass difference of the right-handed
neutrinos is large enough compared with their decay widths, so that the
perturbative calculation is ensured. (We will justify this assumption
later.)

In the present model, the masses $M_{Ri}$ are almost degenerate, $M_{Ri}
= \{{\cal{M}}_R(1-\kappa_R), {\cal{M}}_R(1-\kappa_R),
{\cal{M}}_R(1+2\kappa_R)\}$. Thus, $x\simeq 1$ and the self-energy
contribution ${\cal{F}}_{S}(x)$ is much larger than the
vertex contribution ${\cal{F}}_{V}(x)$.\footnote{If we take into account
the effect of the finite decay width, ${\cal F}_S(x)$ vanishes for $x\to
1$. Thus, ${\cal F}_S(M_k^2/M_i^2)$ in Eq.~(\ref{EQ-epsilon_i}) 
vanish for $i=1,k=2$ and $i=2,k=1$. See also
footnote~\ref{foot-degenerate}.} In the leading order in perturbation,
the asymmetry parameters are given by
\begin{eqnarray}
 \epsilon_1
  =
  \frac{1}{12\pi}
  k_D^2
  \frac{\Im(\kappa_D)}{\Re(\kappa_R)}
  \eps_D^2
  \,,\quad
  \epsilon_2
  =
  \frac{1}{9\pi}
  k_D^2
  \frac{\Im(\kappa_D)}{\Re(\kappa_R)}
  \delta_D^2
  \,,\quad
  \epsilon_3 = \epsilon_1 + \epsilon_2
  \,,
  \label{EQ-epsilons}
\end{eqnarray}
where we have neglected higher order terms in the expansions of
$\eps_D$, $\delta_D$, $\kappa_D$, $\kappa_R$ and $\eps_D/\delta_D$. As
for the $CP$ phase, we have assumed that the complex phase exists only
in the off-diagonal elements $\kappa_D$ and $\kappa_R$ for simplicity,
and have taken the other parameters to be real. Notice that all the
decays of $N_{Ri}$ generate the lepton asymmetry with the same sign,
namely, they contribute in a constructive way.\footnote{One might wonder
if this fact might conflict with the argument that the generated lepton
asymmetry must vanish in the limit of exactly degenerate
masses. However, if the mass differences of the right-handed neutrinos
become smaller than the decay widths of them, the perturbative formula
${\cal F}_S(x)$ in Eq.~(\ref{EQ-fs}) no longer holds and we should take
into account the effect of finite widths of $N_{Ri}$.  Actually, it was
shown that ${\cal F}_S(x)$ vanishes in the limit of exactly degenerate
right-handed neutrino masses if the finite decay widths are
appropriately taken into
account~\cite{degenerate}.
\label{foot-degenerate}}

The ratio of the lepton number density $n_L$ to the entropy density $s$
produced by the inflaton decay is given by~\cite{lep-inf}
\begin{eqnarray}
 \frac{n_L}{s} = \frac{3}{2}\sum_i
  \epsilon_i
  B_r^{(i)}
  \frac{T_R}{m_{\phi}}
  \,,
\end{eqnarray}
where $T_R$ is the reheating temperature after the inflation, $m_{\phi}$
the mass of the inflaton, and $B_r^{(i)}$ the branching ratio of the
decay channel of the inflaton to $N_{Ri}$, i.e., $B_r^{(i)} =
B_r(\phi\to N_{Ri}N_{Ri})$. Here, we have assumed that the inflaton
decays into a pair of right-handed neutrinos, and $M_{Ri}>T_R$ in order
to make the generated lepton asymmetry not washed out by lepton-number
violating processes after the $N_{Ri}$'s decay. Notice that the inflaton
mass $m_\phi$ should satisfy $m_\phi > 2 {\cal{M}}_R$ in order to make
the decay ($\phi\to N_{Ri}N_{Ri}$) kinematically allowed.\footnote{In
this letter, we assume a perturbative decay of the inflaton.} After
being produced, a part of the lepton asymmetry is immediately
converted~\cite{FY} into the baryon asymmetry via the ``sphaleron''
effect~\cite{sphaleron}, since the decays of $N_{Ri}$ take place much
before the electroweak phase transition:
\begin{eqnarray}
 \frac{n_B}{s} = C\frac{n_L}{s}
  \,,
\end{eqnarray}
where $C$ is given by $C \simeq - 0.35$ in the minimal SUSY standard
model (MSSM)~\cite{LtoB}.

Therefore, the amount of the baryon asymmetry in the present model is
estimated as
\begin{eqnarray}
 \frac{n_B}{s}
  &\simeq&
  0.3\times 10^{-10}
  \times
  \left(
   B_r^{(2)} + B_r^{(3)}
   \right)
   \left(\frac{T_R}{10^8\GEV}\right)
  \left(\frac{2 {\cal{M}}_R}{m_\phi}\right)
  \left(\frac{{\cal{M}}_\nu}{0.1\EV}\right)
  \left(\frac{\delta_D}{0.1}\right)^2
  \frac{\Im (\kappa_D)}{\Re (\kappa_R)}
  \,,
  \nonumber\\
 \label{EQ-nBs}
\end{eqnarray}
where we have used the see-saw formula ${\cal{M}}_\nu\simeq k_D^2
\vev{H}^2/{\cal{M}}_R$.\footnote{In the MSSM, $\vev{H} = 174\GEV\times
\sin\beta$, where $\tan\beta = \vev{H}/\vev{H'}$ and $H'$ is the Higgs
field which couples to the down-type quarks (and charged leptons). In
Eq.~(\ref{EQ-nBs}), we have taken $\sin\beta\simeq 1$.} We see that the
empirical baryon asymmetry $n_B/s \simeq (0.4$--$1)\times
10^{-10}$~\cite{PDB} is obtained with a reheating temperature $T_R\simeq
10^8$--$10^{10}\GEV$, for a natural choice of the parameters, say,
$B_r^{(2)}+B_r^{(3)}\simeq0.1$--$1$, $2{\cal M}_R\simeq(0.1$--$1)m_{\phi}$,
$\delta_D\simeq 0.1$ and $\Im (\kappa_D)/\Re (\kappa_R) \simeq 1$.

Notice that the mass differences of the right-handed neutrinos, $\Delta
M_R \simeq 3 \Re(\kappa_R) {\cal M}_R$ is much larger than the decay
widths of them, $\Gamma_{N_{Ri}} \simeq (k_D^2/4\pi) {\cal M}_R$, as
long as $\kappa_R\gg k_D^2/12\pi\simeq 10^{-6}\times ({\cal
M_\nu}/0.1\EV)({\cal M}_R/10^{10}\GEV)$, so that the perturbative
formula ${\cal F}_{S}(x)$ in Eq.~(\ref{EQ-fs}) is justified. The
assumption $M_{Ri} > T_R$ is also easily satisfied for
$M_{Ri}>10^{10}\GEV$.

\begin{figure}[h!]
\begin{center}
\centerline{\psfig{figure=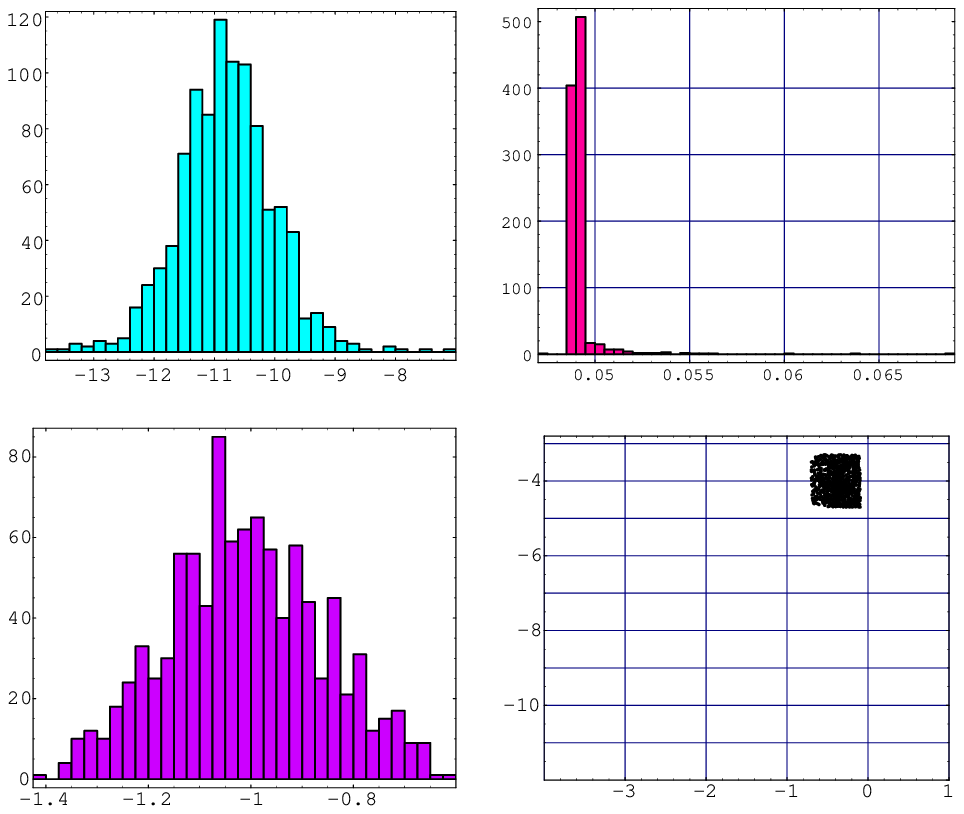,width=18cm}}
\begin{picture}(0,0)
\large
\put(-145,233){${\rm{log}}_{10}[n_{B}/s]$}
\put(100,230){${\rm{log}}_{10}\left[|\rm{U}_{e3}|\right]$}
\normalsize
\put(-150,10){${\rm{log}}_{10}\left[\displaystyle{\frac{{\cal{M}}_{\nu}}{\EV}}
\right]$}\large
\put(105,15){${\rm{log}}_{10}[{\rm{tan}}^{2}\theta_{12}]$}
\normalsize
\put(-6,130){$\Delta m^{2}_{\rm{sol}}$}
\end{picture} 
\end{center}
\caption{Histograms of the $n_{B}/s$, $|{\rm{U}}_{e3}|$ and
 ${\cal{M}}_{\nu}$. We randomly generate the small perturbations as in
 Eq.~(\ref{range-sol}), and collect the data set of these quantities if
 the generated mass spectrum and mixing angles of the light neutrinos
 satisfy the conditions for the LMA solution in Eq.(\ref{experiment}).
 Then we show the frequencies of these quantities in the vertical axes.
 The generated mass spectrum and mixing angles for the solar neutrino
 oscillation is also plotted in the lower-right figure, in which $\Delta
 m^{2}_{\rm{solar}}$ is plotted with ${\rm{log}}_{10}[\Delta
 m^{2}_{\rm{solar}}/\EV^{2}]$ unit. } \label{fig:Ftimes5Mun}
\end{figure}

In Fig.~\ref{fig:Ftimes5Mun}, we show the histograms of the $n_{B}/s$,
$|{\rm{U}}_{e3}|$ and ${\cal{M}}_{\nu}$ for the LMA solution. Here, we
randomly generate the small perturbations as in Eq.~(\ref{perturbation})
and collect the data set of $n_B/s$, $|{\rm{U}}_{e3}|$ and
${\cal{M}}_{\nu}$ if the generated mass spectrum and mixing angles
satisfy the conditions obtained from the neutrino oscillation
experiments given in Eq.~(\ref{experiment}). Then we plot the
frequencies of these quantities in the vertical axes.  We also present a
plot of the obtained mass differences and mixing angles for the solar
neutrino oscillation in $\{{\rm{tan}}^2 \theta_{12},\Delta
m^{2}_{\rm{solar}}\}$ plane.  We take the range of the perturbations as
follows:
\begin{eqnarray}
&&\delta_{D}=(0.25-1.5)\times 10^{-1},\quad\eps_{D}= \delta_{D}10^{-(0.3-2.5)},
\nonumber\\
&&\kappa_{D}=\eps_{D}10^{-(0-0.7)}e^{i (0-2)\pi },\quad
\kappa_{R}=|\kappa_{D}|e^{i(-\frac{1}{2}\sim\frac{1}{2})\pi},
\label{range-sol}
\end{eqnarray}
where we take $\eps_{D}$ and $\delta_{D}$ to be real, for simplicity.
If we randomly generate the perturbations within the above range,
the neutrino mass spectrum and mixing angles satisfy the conditions
given in Eq.~(\ref{experiment}) with the rate of nearly 
$20\%$. 
The baryon asymmetry $n_{B}/s$ is calculated with the  normalization 
\begin{equation}
 (B_{r}^{(2)}+B_{r}^{(3)})\left(\frac{T_{R}}{10^{8}\GEV}\right)
  \left(\frac{2{\cal{M}}_{R}}{m_{\phi}}\right)=1\,.
\end{equation} 
One can see that the analytic estimation explains the numerical result
quite well and that the required amount of the baryon asymmetry can be
easily obtained in the democratic model with the natural scale of
perturbations. Another interesting prediction can be seen from the
upper-right figure.  The amplitude of U$_{e3}$ is accurately predicted
as $|{\rm{U}}_{e3}|\simeq0.049$, as denoted in the first part of this
letter.

\begin{figure}[h!]
\begin{center}
\centerline{\psfig{figure=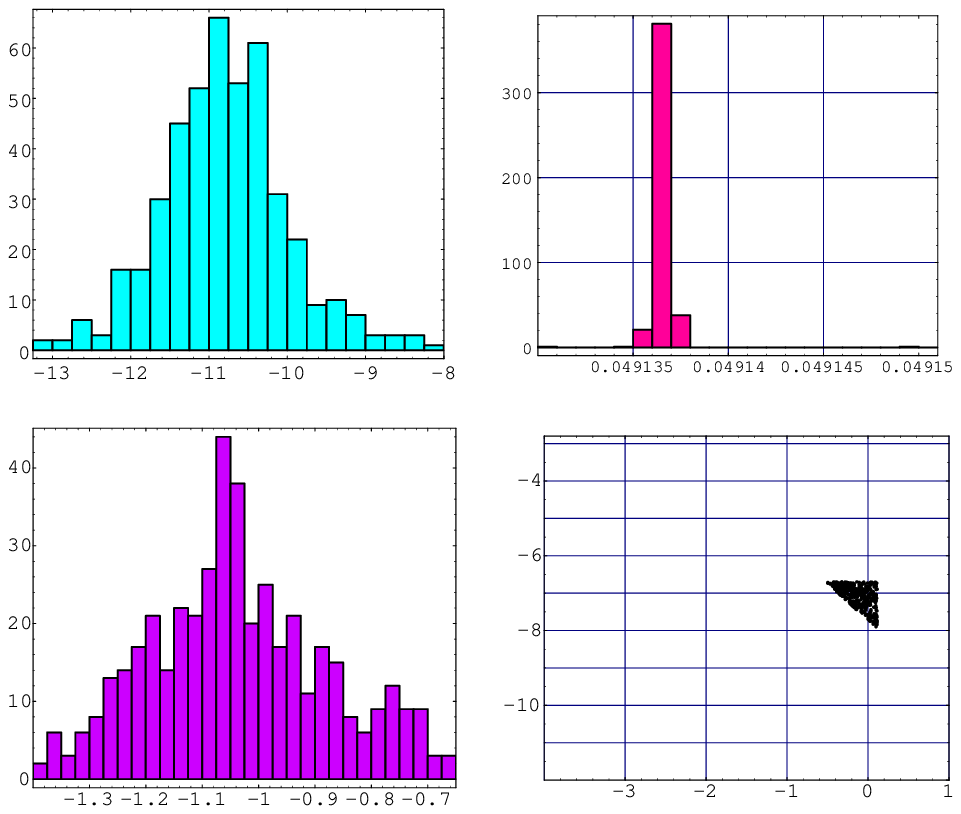,width=18cm}}
\begin{picture}(0,0)
\large
\put(-145,233){${\rm{log}}_{10}[n_{B}/s]$}
\put(100,230){${\rm{log}}_{10}\left[|\rm{U}_{e3}|\right]$}
\normalsize
\put(-150,10){${\rm{log}}_{10}\left[\displaystyle{\frac{{\cal{M}}_{\nu}}{\EV}}
\right]$}\large
\put(105,15){${\rm{log}}_{10}[{\rm{tan}}^{2}\theta_{12}]$}
\normalsize
\put(-9,130){$\Delta m^{2}_{\rm{solar}}$}
\end{picture} 
\end{center}
\caption{The same as Fig.{\ref{fig:Ftimes5Mun}}, but for the LOW solution.
}
 \label{fig:low}
\end{figure}

In Fig.~\ref{fig:low}, we show the histogram of the the
$n_{B}/s$, $|{\rm{U}}_{e3}|$ and ${\cal{M}}_{\nu}$ for  the LOW solution.
In this case, we take the range of perturbations as follows:
\begin{eqnarray}
&&\delta_{D}=(0.25-1.5)\times 10^{-1},\quad\eps_{D}= \delta_{D}10^{-(3.7-5.8)},
\nonumber\\
&&\kappa_{D}=\eps_{D}10^{-(0-0.7)}e^{i (0-2)\pi },\quad
\kappa_{R}=|\kappa_{D}|e^{i(-\frac{1}{2}\sim\frac{1}{2})\pi}.
\label{range-low}
\end{eqnarray}
Other conventions are the same as those in Fig~\ref{fig:Ftimes5Mun}.
The resultant baryon asymmetry is almost the same 
as in the LMA solution, which is easily understood from the
Eq.~(\ref{EQ-nBs}). In the case of the LOW solution,
the off diagonal elements in the light
neutrino mass matrix are much smaller than those in the LMA 
solution $\kappa_{\nu}\sim \eps_{\nu}\sim \delta_{\nu}\times 10^{-(4-6)}$, which results in the much more precise prediction on the 
amplitude of the U$_{e3}$.

{\it Conclusions}

In this letter we have investigated leptogenesis with almost degenerate
neutrinos, in the framework of democratic mass matrix, which explains
very successfully the observed large mixings of the neutrinos as well as
quark masses and mixings. The almost degenerate Majorana neutrinos with
masses of order $m_i^\nu\sim {\cal O}(0.1)\EV$ induce a considerable rate of
the neutrinoless double beta decays, which is accessible in near future
experiments.  We have shown that the empirical baryon asymmetry is well
explained by the decays of the right-handed neutrinos produced in the
inflaton decay.

In this model, the U$_{e3}$ component of the mixing matrix for neutrino
oscillations is predicted as $|{\rm U}_{e3}|\simeq 0.05$, which is a
direct consequence of the nearly S$_3$ symmetric form of the neutrino
mass matrix. Such a value of U$_{e3}$ will be also tested in the long
baseline experiments, such as JHF~\cite{JHF}.

{\it Appendix}

In this appendix, we show that if the light neutrinos are degenerate
in masses of $m_{i}^\nu\simeq {\cal O}(0.1)\EV$, the produced lepton
asymmetry is highly suppressed in the case of the leptogenesis scenario
where the right-handed neutrinos are produced by thermal scatterings.
We should stress that the following discussion is a generic consequence
of the almost degenerate light neutrinos, independent of the specific models.

The wash-out effect of the lepton asymmetry in the leptogenesis by the
decays of thermally produced right-handed neutrino $N_{Ri}$ crucially
depends on the following parameter $K_i$:
\begin{eqnarray}
 K_i\equiv \frac{\Gamma_{N_{Ri}}}{H(T = M_{Ri})}
  \,,
\end{eqnarray}
where $H(T=M_{Ri})$ is the Hubble parameter when the temperature $T$
becomes as low as the mass of the decaying right-handed neutrino,
$M_{Ri}$. The so-called out-of-equilibrium condition is roughly given by
$K_i\lsim 1$, and the final lepton asymmetry is strongly suppressed for
$K_i \gg 1$~\cite{Buchmuller-Plumacher}. The parameter $K_i$ can be
rewritten in terms of a mass parameter $\widetilde{m_i}$~\cite{Buchmuller-Plumacher}:
\begin{eqnarray}
 K_i \simeq \frac{\widetilde{m_i}}{0.001\EV}
  \,,
  \label{EQ-K-mtilde}
\end{eqnarray}
where
\begin{eqnarray}
 \widetilde{m_i}
  \equiv
  \sum_k
  |\widetilde{h}_{ik}|^2
  \frac{\vev{H}}{M_{Ri}}
  \,.
  \label{EQ-mitilde}
\end{eqnarray}
We use the Yukawa couplings $\widetilde{h}_{ik}$ defined in the basis
where the both mass matrices of right-handed neutrinos and light
neutrinos are diagonal:
\begin{eqnarray}
 \sum_i \widetilde{h}_{ij} \widetilde{h}_{ik}
  \frac{\vev{H}}{M_{Ri}}
  =
  m_j^\nu \delta_{jk}
  \,.
  \label{EQ-XXT}
\end{eqnarray}
Let us define a matrix $X_{ij}$ as follows:
\begin{eqnarray}
 X_{ij}\equiv 
  \widetilde{h}_{ij}
  \frac{\vev{H}}{\sqrt{M_{Ri}\,m_j^\nu}}
  \,.
\end{eqnarray}
Then from Eq.~(\ref{EQ-XXT}) one can show
\begin{eqnarray}
 \left(X^T X\right)_{ij} = \delta_{ij} = \left(X X^T\right)_{ij}
  \,.
  \label{EQ-XXT-1}
\end{eqnarray}
Thus, we see that the $\widetilde{m_i}$ parameter in
Eq.~(\ref{EQ-mitilde}) is bounded from below as follows:
\begin{eqnarray}
 \widetilde{m_i}
  &=&
  \sum_k
  m_k^\nu
  \left|X_{ik} \right|^2
  \nonumber\\
 &>&
  \min_j \{m_j^\nu\}
  \sum_k
  \left|X_{ik}\right|^2
  \nonumber\\
 &\ge&
  \min_j \{m_j^\nu\}
  \left|\sum_k
  X_{ik}^2\right|
  \nonumber\\
 &=&
  \min_j \{m_j^\nu\}
  \,.
\end{eqnarray}
In the last equation, we have used Eq.~(\ref{EQ-XXT-1}). Therefore, if
the light neutrinos are degenerate as $m_1^\nu\simeq m_2^\nu\simeq
m_3^\nu\simeq {\cal O}(0.1)\EV$, the $K$ parameter becomes at least as
large as $K > {\cal O}(100)$ [see Eq.~(\ref{EQ-K-mtilde})], which leads
to a strong suppression of the lepton asymmetry generated by the decays
of thermally produced right-handed neutrinos.


\newpage

\begin{thebibliography}{99}

 \bibitem{Harari-etal}
 H.~Harari, H.~Haut and J.~Weyers,
 Phys.\ Lett.\ B {\bf 78} (1978) 459.

 \bibitem{Koide}
 Y.~Koide,
 Phys.\ Rev.\ D {\bf 28} (1983) 252;
 Phys.\ Rev.\ D {\bf 39} (1989) 1391.

 \bibitem{FTY}
 M.~Fukugita, M.~Tanimoto and T.~Yanagida,
 Phys.\ Rev.\ D {\bf 57} (1998) 4429
 [arXiv:hep-ph/9709388].

 \bibitem{Fritzsch-etal-2}
 H.~Fritzsch and Z.~Z.~Xing,
 Phys.\ Lett.\ B {\bf 440} (1998) 313
 [arXiv:hep-ph/9808272].


 \bibitem{Fritzsch-etal-1}
 H.~Fritzsch and Z.~Z.~Xing,
 Phys.\ Lett.\ B {\bf 372} (1996) 265
 [arXiv:hep-ph/9509389].


 \bibitem{CHOOZ}
 M.~Apollonio {\it et al.},
 Phys.\ Lett.\ {\bf B466} (1999) 415
 [arXiv:hep-ex/9907037].


 \bibitem{SK-Atm}
 Y.~Fukuda {\it et al.}  [Super-Kamiokande Collaboration],
 \\
 Phys.\ Lett.\ B {\bf 433} (1998) 9
 [arXiv:hep-ex/9803006];
 \\
 Phys.\ Lett.\ B {\bf 436} (1998) 33
 [arXiv:hep-ex/9805006];
 \\
 Phys.\ Rev.\ Lett.\  {\bf 81} (1998) 1562
 [arXiv:hep-ex/9807003].
 \\
 See also a recent data,
 \\
 C.~Yanagisawa,
 Nucl.\ Phys.\ Proc.\ Suppl.\  {\bf 95} (2001) 93.
 %



 \bibitem{SK-solar}
 Y.~Fukuda {\it et al.}  [Super-Kamiokande Collaboration],
 \\
 Phys.\ Rev.\ Lett.\  {\bf 81} (1998) 1158
 [Erratum-ibid.\  {\bf 81} (1998) 4279]
 [arXiv:hep-ex/9805021];
 \\
 Phys.\ Rev.\ Lett.\  {\bf 82} (1999) 2430
 [arXiv:hep-ex/9812011];
 \\
 Phys.\ Rev.\ Lett.\  {\bf 82} (1999) 1810
 [arXiv:hep-ex/9812009].

 

 \bibitem{GENIUS}
 H.~V.~Klapdor-Kleingrothaus {\it et al.}  [GENIUS Collaboration],
 arXiv:hep-ph/9910205.

 \bibitem{CUORE}
 E.~Fiorini,
 Phys.\ Rept.\  {\bf 307} (1998) 309;
 Nucl.\ Phys.\ Proc.\ Suppl.\  {\bf 100} (2001) 332.

 \bibitem{MOON}
 H.~Ejiri, J.~Engel, R.~Hazama, P.~Krastev, N.~Kudomi and R.~G.~Robertson,
 Phys.\ Rev.\ Lett.\  {\bf 85} (2000) 2917
 [arXiv:nucl-ex/9911008].


 \bibitem{XMASS}
 S.~Moriyama, Talk at International Workshop on Technology 
 and Application of Xenon Detectors (Xenon01), ICRR, Kashiwa, Japan, 
 December 3-4, 2001.

 \bibitem{EXO}
 S.~Waldman, Talk at International Workshop on Technology
 and Application of Xenon Detectors (Xenon01), ICRR, Kashiwa, Japan,
 December 3-4, 2001.
 













 \bibitem{FY}
 M.~Fukugita and T.~Yanagida,
 Phys.\ Lett.\ {\bf B174} (1986) 45.

 \bibitem{lep-inf}
 K.~Kumekawa, T.~Moroi and T.~Yanagida,
 Prog.\ Theor.\ Phys.\  {\bf 92} (1994) 437
 [arXiv:hep-ph/9405337];
 \\
 G.~Lazarides,
 Springer Tracts Mod.\ Phys.\  {\bf 163} (2000) 227
 [arXiv:hep-ph/9904428]
 and references therein;
 \\
 G.~F.~Giudice, M.~Peloso, A.~Riotto and I.~Tkachev,
 JHEP {\bf 9908} (1999) 014
 [arXiv:hep-ph/9905242];
 \\
 T.~Asaka, K.~Hamaguchi, M.~Kawasaki and T.~Yanagida,
 Phys.\ Lett.\ B {\bf 464} (1999) 12
 [arXiv:hep-ph/9906366],
 Phys.\ Rev.\ D {\bf 61} (2000) 083512
 [arXiv:hep-ph/9907559].

 \bibitem{postSNO}
 G.~L.~Fogli, E.~Lisi, D.~Montanino and A.~Palazzo,
 Phys.\ Rev.\ D {\bf 64} (2001) 093007
 [arXiv:hep-ph/0106247].
 \\
 G.~L.~Fogli, E.~Lisi, A.~Marrone, D.~Montanino and A.~Palazzo,
 arXiv:hep-ph/0201290,
 and references therein.

 \bibitem{H-M-0nubb}
 H.~V.~Klapdor-Kleingrothaus, A.~Dietz, H.~L.~Harney and I.~V.~Krivosheina,
 Mod.\ Phys.\ Lett.\ A {\bf 16} (2002) 2409
 [arXiv:hep-ph/0201231].

 \bibitem{JHF}
 Y.~Itow {\it et al.},
 arXiv:hep-ex/0106019.
 \\
 See also \verb|http://neutrino.kek.jp/jhfnu/|.

 \bibitem{seesaw}
 T.~Yanagida,
 in Proceedings of the 
 {\it ``Workshop on the Unified Theory and the Baryon Number in the
 Universe''}, Tsukuba, Japan, 1979, 
 edited by O.~Sawada and A.~Sugamoto, KEK Report No. KEK-79-18, p. 95;
 Prog.\ Theor.\ Phys.\  {\bf 64} (1980) 1103;
 \\
 M.~Gell-Mann, P.~Ramond and R.~Slansky,
 in {\it ``Supergravity''},
 edited by D.Z. Freedman and P. van Nieuwenhuizen
 (North-Holland, Amsterdam, 1979).
 

 \bibitem{Buchmuller-Plumacher}
 For reviews and references, see, for example, \\
 M.~Plumacher,
 Nucl.\ Phys.\ B {\bf 530} (1998) 207
 [arXiv:hep-ph/9704231];
 \\
 W.~Buchmuller and M.~Plumacher,
 Phys.\ Rept.\  {\bf 320} (1999) 329
 [arXiv:hep-ph/9904310];
 \\
 W.~Buchmuller and M.~Plumacher,
 Int.\ J.\ Mod.\ Phys.\ A {\bf 15} (2000) 5047
 [arXiv:hep-ph/0007176].
 



 \bibitem{epsilon1}
 M.~Flanz, E.~A.~Paschos and U.~Sarkar,
 Phys.\ Lett.\ B {\bf 345} (1995) 248
 [Erratum-ibid.\ B {\bf 384} (1995) 487]
 [arXiv:hep-ph/9411366];
 \\
 L.~Covi, E.~Roulet and F.~Vissani,
 Phys.\ Lett.\ B {\bf 384} (1996) 169
 [arXiv:hep-ph/9605319];
 \\
 W.~Buchmuller and M.~Plumacher,
 Phys.\ Lett.\ B {\bf 431} (1998) 354
 [arXiv:hep-ph/9710460].

 \bibitem{epsilon1-SUSY}
 L.~Covi, E.~Roulet and F.~Vissani, in Ref.~\cite{epsilon1}.

 \bibitem{degenerate}
 For a discussion and references, see A.~Pilaftsis,
 Int.\ J.\ Mod.\ Phys.\ A {\bf 14} (1999) 1811
 [arXiv:hep-ph/9812256].


 \bibitem{sphaleron}
 V.~A.~Kuzmin, V.~A.~Rubakov and M.~E.~Shaposhnikov,
 Phys.\ Lett.\ B {\bf 155} (1985) 36.


 \bibitem{LtoB}
 S.~Y.~Khlebnikov and M.~E.~Shaposhnikov,
 Nucl.\ Phys.\ {\bf B308} (1988) 885;
 \\
 J.~A.~Harvey and M.~S.~Turner,
 Phys.\ Rev.\ {\bf D 42} (1990) 3344.

 \bibitem{PDB}
 D.~E.~Groom {\it et al.}  [Particle Data Group Collaboration],
 ``Review of particle physics,''
 Eur.\ Phys.\ J.\ C {\bf 15} (2000) 1.




\end{thebibliography}
\end{document}